\documentclass[letterpaper]{article} 
\usepackage{aaai25}  
\usepackage{times}  
\usepackage{helvet}  
\usepackage{courier}  
\usepackage[hyphens]{url}  
\usepackage{graphicx} 
\usepackage{natbib}  
\usepackage{caption} 
\usepackage{algorithm}
\usepackage{algorithmic}

\usepackage{xcolor}
\usepackage[most]{tcolorbox}
\usepackage{longtable}

\usepackage{newfloat}
\usepackage{listings}
\DeclareCaptionStyle{ruled}{labelfont=normalfont,labelsep=colon,strut=off} 
\floatstyle{ruled}
\newfloat{listing}{tb}{lst}{}
\floatname{listing}{Listing}
\usepackage{subcaption}
\usepackage{xr}
\usepackage{booktabs}

\title{Motivation, Attention, and Visual Platform Design: How Moral Contagions Spread on TikTok and Instagram in the 2024 United States Presidential Election}
\author{
    Ni Annie Yuan$^1$,
    Ho-Chun Herbert Chang$^1$
}
\affiliations{
    \textsuperscript{\rm 1} Program in Quantitative Social Science\\

    Dartmouth College\\
    Hanover, NH 03755 USA\\
    annie.yuan.28@dartmouth.edu
}

\begin{document}

\maketitle

\begin{abstract}
Visual social media platforms have become primary venues for political discourse, yet we know little about how moralization operates differently across platforms and topics. Analyzing 2,027,595 TikToks and 1,126,972 Instagram posts during the 2024 US presidential election, we demonstrate that issues are not necessarily inherently moralized, but a product of audience demographics, platform architecture, and partisan framing. Using temporal supply-demand analysis and moral foundations scoring (eMFD), we examine the dynamics of key electoral issues. Three key findings emerge. First, moralization patterns diverge dramatically by platform: TikTok's algorithm enabled viral spread of moralized abortion and immigration content despite lower supply, while Instagram amplified economic discourse that aligned supply and demand. Second, traditionally "pragmatic" economic issues became moralized—cryptocurrency discourse invoked loyalty and authority foundations more strongly than any other topic, framing regulation as government overreach. Third, platforms responded to different events: TikTok surged after Harris's nomination across all topics (96\% reduction in supply volatility), while Instagram spiked around cryptocurrency policy developments. Semantic network analysis reveals TikTok's circular topology enables cross-cutting exposure while Instagram's fragmented structure isolates Harris from economic discourse. These findings demonstrate that understanding political moralization requires examining platform-specific ecosystems where architecture, demographics, and content strategy interact to determine which issues get moralized and how moral content spreads.
\end{abstract}

%

\section{Introduction}


The 2024 US presidential election unfolded during a period of growing ideological polarization and increased social media engagement among American voters~\cite{Brenan2025GallupIdeologicalPolarization}, with Donald Trump representing the Republican Party and Kamala Harris the incumbent Democrat Party.  Crucially, visual-first social media platforms—TikTok and Instagram—have in recent years rivaled or exceeded text-based networks like Twitter/X as primary venues for political engagement, particularly among young voters \cite{9}.

This shift to visual platforms carries profound implications for political persuasion~\cite{ChangRichardsonFerrara2022}. Research on "moral contagions"---content that frames political issues in terms of right and wrong, good and evil—has consistently found that moralized messages spread faster and wider than neutral information, particularly when they evoke strong emotions like anger and disgust~\cite{8,7}. Visual media, with its capacity to provoke immediate emotional responses, should theoretically accelerate this process. Yet few studies have directly compared moralization across different platforms like TikTok and Instagram.

This study addresses a core puzzle: when the same political issues are discussed on different visual platforms, do moralization patterns diverge? 

We focus on three issues central to the 2024 election: abortion, immigration, and the economy. Abortion emerged as the top priority for liberal voters following the overturning of Roe v. Wade, while the economy and immigration dominated conservative concerns~\cite{Brenan2025GallupIdeologicalPolarization,19}. These follow longer trends of issue salience by the two parties~\cite{kunda1990case}. Abortion and immigration have long been moralized in American political discourse, framed in terms of fundamental rights, national identity, and moral duty. The economy, by contrast, typically generates pragmatic rather than moralized debate. This variation creates an opportunity: if visual platforms uniformly intensify moralization, we should observe all three issues receiving moral framing. If platforms shape moralization differently, we might find selective patterns—some issues moralized on one platform but not another.

Analyzing 2,027,595 TikToks and 1,126,972 Instagram posts from January through November 2024, we uncover striking platform divergences. Relevant categories were analyzed in terms of supply (raw volume) and demand (views amassed) over time~\cite{munger2022right,ZhaChang2025GenderInequalities}. TikTok's algorithm, which prioritizes novel content regardless of creator following, enabled viral spread of moralized content addressing abortion and immigration. Instagram's hybrid model favored economic policy discussions—which our moral foundations analysis reveals were themselves moralized through loyalty and authority framings around government overreach and cryptocurrency regulation. Crucially, our analysis allows us to probe the three elements that contribute to political discourse--- motivation through moralized issue-framing, attention through temporal supply-demand dynamics, and design through comparison of TikTok and Instagram.

Our findings challenge the assumption that visual platforms uniformly accelerate moral contagion. Instead, we demonstrate that algorithmic design and demographic composition create distinct moral-political ecosystems. Understanding these differences matters urgently: as visual platforms become primary sites of political persuasion, the mechanisms by which they moralize—or fail to moralize—political issues will shape democratic discourse in consequential ways.

\section{Related Works}

\subsection{Moralization of Political Discourse Online}

 American political discourse has become increasingly moralized: more issues are framed as matters of fundamental right and wrong rather than pragmatic trade-offs \cite{8}. This shift has accelerated dramatically on social media platforms, where moral language use on Twitter and Reddit increased between 2012 and 2023 at rates exceeding those in traditional media \citep{puryear2025rising}. Understanding why moralization spreads rapidly online—and how platform-specific features shape this process—has become critical to understanding contemporary political polarization.

 The concept of ``moral contagion" describes how moralized political content spreads through social networks via emotional engagement and social reinforcement \citep{8}. Moral framing—connecting political issues to fundamental values about right and wrong—captures attention more effectively than neutral presentation. When content evokes ``other-condemning" emotions like anger and disgust, it triggers both moral conviction in viewers and motivation to share the content with others \citep{7, haidt2003moral}. This creates a feedback loop: exposure to moralized content strengthens users' moral beliefs about issues, increasing their likelihood of engaging with and spreading similar content, which then reaches broader audiences. The psychological rewards of moral signaling, combined with algorithmic amplification of high-engagement content, create conditions for rapid spread of moral framings.

Empirical evidence supports this theoretical model. \citet{valenzuela2017behavioral} found that morally-framed news content on social media generated significantly more sharing than content framed in economic or human-interest terms. \citet{7} demonstrated experimentally that exposure to emotionally moralized frames increased participants' moral conviction about political issues, with effects persisting at least two weeks. Importantly, these effects occurred across the political spectrum—moral framing increased conviction among both liberals and conservatives, though the specific moral foundations invoked (care vs. loyalty, fairness vs. authority) varied by ideology \cite{graham2009liberals}.

\subsection{Operationalization and Moral Foundations}

Moral Foundations Theory (MFT) posits that moral judgments derive from five innate psychological foundations: care/harm, fairness/cheating, loyalty/betrayal, authority/subversion, and sanctity/degradation \citep{haidt2012righteous, graham2013moral}. Critically, these foundations are not equally salient across the political spectrum. Liberals prioritize individualizing foundations (care and fairness), emphasizing harm prevention and equal treatment, while conservatives weight all five foundations more evenly, with particular emphasis on binding foundations (loyalty, authority, sanctity) that reinforce group cohesion and traditional hierarchies \citep{graham2009liberals, graham2011mapping}.

This ideological asymmetry has profound implications for political communication. Messages invoking foundations aligned with an audience's moral psychology should resonate more strongly and spread more readily. For instance, pro-choice abortion rhetoric emphasizing bodily autonomy and preventing harm to women invokes care and fairness foundations, while pro-life messaging about the sanctity of life and protecting innocent beings draws on sanctity and care from a different angle \citep{clifford2015moral}. Immigration discourse similarly divides along foundation lines: arguments about humanitarian obligations invoke care, while concerns about national sovereignty and cultural preservation invoke loyalty and authority \citep{koleva2012tracing}.

Early applications of MFT to political discourse relied on manual coding, limiting scalability. The development of the Moral Foundations Dictionary (MFD) enabled automated text analysis by mapping words to moral foundations \citep{graham2009liberals}. However, the original MFD suffered from limited lexical coverage and inability to handle context-dependent meanings. The extended Moral Foundations Dictionary (eMFD) addressed these limitations by employing probabilistic word-foundation associations rather than binary classifications, and by incorporating a much larger vocabulary learned from human annotations of moral content \citep{hopp2021extended}.

\citet{hopp2021extended} validated eMFD through multiple studies, demonstrating that it outperforms the original MFD in predicting human moral judgments and exhibits stronger correlations with political ideology. Importantly, eMFD captures both the presence and intensity of moral language, enabling fine-grained analysis of how strongly content invokes particular foundations. This methodological advance has enabled large-scale studies of moral framing across social media platforms.

Recent applications of eMFD to political social media have yielded key insights. \citet{7} found that moral-emotional language in political tweets predicted both within-network and cross-network sharing, with each additional moral-emotional word increasing retweet probability by 20\%. \citet{mooijman2018moralization} used MFD to show that moral framing of climate change increased among both liberals and conservatives between 2008 and 2017, though the specific foundations invoked diverged by ideology. \citet{voelkel2022interventions} demonstrated that highlighting moral common ground across foundations could reduce political animosity, suggesting that understanding foundation-specific appeals matters for both polarization and depolarization efforts.

However, existing eMFD applications have focused almost exclusively on text-based platforms like Twitter. Visual platforms present unique challenges: captions and descriptions are often shorter and less linguistically rich than tweets, while the visual content itself may carry moral signals not captured by text analysis alone. Moreover, no studies have systematically compared moral foundation invocation across platforms with different algorithmic architectures. If TikTok's content-driven algorithm versus Instagram's network-driven model shapes which content goes viral, we might expect platform-specific patterns in which moral foundations prove most effective for achieving reach.

\subsection{Issue-Specific Framing and Moralization}

Different political issues naturally align with different moral foundations, shaping how they're discussed and who they persuade. Abortion discourse provides perhaps the clearest example of foundation-based framing divergence. \citet{clifford2015moral} content-analyzed abortion arguments and found that pro-choice frames predominantly invoked care (for women's wellbeing) and fairness (reproductive rights, bodily autonomy), while pro-life frames emphasized sanctity (sacred value of life) and care reframed toward the fetus. Importantly, exposing participants to foundation-matched frames increased moral conviction more than foundation-mismatched frames, demonstrating that strategic foundation invocation enhances persuasive impact.

Immigration attitudes similarly map onto moral foundations. \citet{koleva2012tracing} found that support for restrictive immigration policies correlated with loyalty (in-group preference), authority (respect for legal order), and sanctity (concerns about cultural purity) foundations, while opposition correlated with care (empathy for immigrants) and fairness (equality of opportunity). \citet{voelkel2019moral} review evidence that reframing political arguments to invoke the moral foundations of the opposing side---such as invoking loyalty and authority when persuading conservatives---is significantly more effective than using one's own moral foundations, including on issues such as immigration. In general, partisan asymmetries in textual social media platforms have been documented~\cite{chang2025liberals}.

On the other hand, traditional economic debates center on efficiency and outcomes rather than moral principles. However, recent work suggests economic issues can be moralized through foundation-specific frames. \citet{voelkel2020economic} demonstrated that high economic inequality is linked to greater moralization across all five moral foundations, suggesting economic conditions can trigger moral framing that extends beyond traditionally moralized issues. The rise of cryptocurrency discourse may represent a similar moralization process, with regulation framed as government overreach (authority/subversion) and financial freedom as an in-group identity marker (loyalty).

\subsection{Platform-Specific Moral Dynamics}

While moral foundations theory explains what moral content communicates, platform architecture shapes which moral content spreads. \citet{8} propose the MAD (Motivation, Attention, Design) model, arguing that platform design features—algorithms, sharing affordances, network structures—fundamentally moderate moral contagion. Their model predicts that design choices determining content visibility will interact with moral content's attention-capturing properties to produce platform-specific virality patterns.

Emerging evidence supports this prediction. \citet{rathje2021outgroup} found that posts expressing out-group animosity gained more engagement on Twitter, with each out-group term increasing shares by 67\%, but this effect varied by users' network structures—those with more ideologically diverse networks showed weaker out-group amplification. \citet{shin2024algorithms} audited TikTok's recommendation algorithm and found that platform recommendations constituted a primary pathway through which users accessed far-right content, suggesting that content-driven algorithms may have lower thresholds for surfacing moralized political material. Separately, \citet{robertson2023users} found that users on Google Search actively chose to engage with more partisan news than the algorithm exposed them to, highlighting the interplay between algorithmic curation and user agency.

\subsection{Research Questions}
Our research aims to analyze several gaps. By applying eMFD to over 3 million posts across TikTok and Instagram during the 2024 election, we thereby examine how content is moralized, but also more granularly, which foundations are invoked, with what intensity, and on what platforms. Our RQs are designed to answer each of Motivation, Attention, Design, and are the following:
\begin{enumerate}
    \item \textbf{RQ1: (Attention):} How do issue-specific supply and demand dynamics change across the two platforms? How do they respond to external political events?
    \item \textbf{RQ2: (Motivation):} Which moral foundations are invoked across issues and candidates, and does this differ by platform?
    \item \textbf{RQ3 (Design):} How does platform architecture structure political discourse networks and candidate positioning?
\end{enumerate}

\section{Data and Methods}
\subsection{Data Collection}
The analysis utilizes data from 2,027,595 relevant TikToks between January 1st and November 14th, as well as 1,126,972 Instagram posts between January 1st and October 28th of 2024. This data was collected using two streams, one for each platform. Instagram data was sourced from Meta Content Library, and the official TikTok Researcher API was used to collect data from TikTok. Posts with politically relevant content were filtered for with snowball sampling using a list of key words that can be found in the Appendix.

\subsection{Content Categorization}
The data for both platforms contained a ``hashtags'' attribute for each post, listing all the hashtags used by the poster in description of said post. This attribute was used to determine the alignment of post content with any of the four topics of interest: pro-life-aligned abortion, pro-choice-aligned abortion, politically neutral economy, and right-leaning immigration. Each topic was defined with a list of relevant hashtags, such as \textit{\#prolife}, \textit{\#pro-life}, \textit{\#prolifegeneration}, \textit{\#banabortion}, \textit{\#abortionismurder}, etc. for pro-life aligned abortion. These hashtags were generated through a process of snowball sampling by identifying frequently used hashtags associated with main hashtags from each category, like \textit{\#prolife}, \textit{\#prochoice}, \textit{\#immigration}, and \textit{\#economy.} Complete lists of hashtags used for categorization can be found in the Appendix. Any post containing at least one relevant tag was deemed as pertaining to the associated category, with each post having the potential to be associated with none or multiple categories.

\subsection{Quantified Supply and Demand}
The supply of content was quantified as the total number of posts attributed to each of the four categories across the two platforms each day. Demand for content across each topic was determined by the average views generated by each post per day. The specific number was calculated as the total views generated by all posts created under each topic on a given day divided by the supply of posts for that day. Both metrics were smoothed across seven-day intervals.

\subsection{Network Visualizations}
In addition to timeseries analysis, we also provide two visualizations of hashtag co-usage across each platform. Hashtag pairs were extracted from all posts labeled with at least one of the four categories of interest and the top 1,000 most frequent pairs were used to populate the networks, which were constructed using the NetworkX package in Python. Each unique hashtag in a list of pairs corresponds to a node in a network, and an edge between nodes represents the corresponding hashtags being used together in one or more posts. These networks were visualized using Netwulf with the same visualization parameters to maintain consistency. 

\subsection{Morality Scoring with eMFD}

We applied the extended Moral Foundations Dictionary (eMFD) to measure moral content across posts \citep{hopp2021extended}. Unlike the original Moral Foundations Dictionary which uses binary word-foundation classifications, eMFD employs probabilistic associations between words and the five moral foundations (care, fairness, loyalty, authority, sanctity). These associations were derived through crowdsourced annotations where participants rated the extent to which words relate to each foundation on a continuous scale. For each post, we extracted the captions. eMFD assigns each word a probability score (0-1) for its association with each foundation, which denote both the strength and direction (virtue vs. vice) of moral content. We calculated document-level foundation scores by aggregating word-level probabilities across all words in a post, following \citet{hopp2021extended}'s recommended approach. This yields a moral foundation profile for each post indicating the relative emphasis on each foundation (package available on GitHub).

To account for emotional valence, we incorporated sentiment weighting: positive sentiment amplifies virtue-related foundation scores (e.g., care as compassion) while negative sentiment amplifies vice-related scores (e.g., care violations as harm). Sentiment was determined using eMFD's built-in sentiment lexicon. Final scores represent the average sentiment-weighted moral foundation strength across all posts within each topic category (pro-life, pro-choice, immigration, economy) and candidate hashtag (\textit{\#trump}, \textit{\#biden}, \textit{\#kamalaharris}).

\section{Results}

\subsection{Supply and demand dynamics of attention}
\begin{figure*}[!htb]
    \centering
    \includegraphics[width=0.9\textwidth]{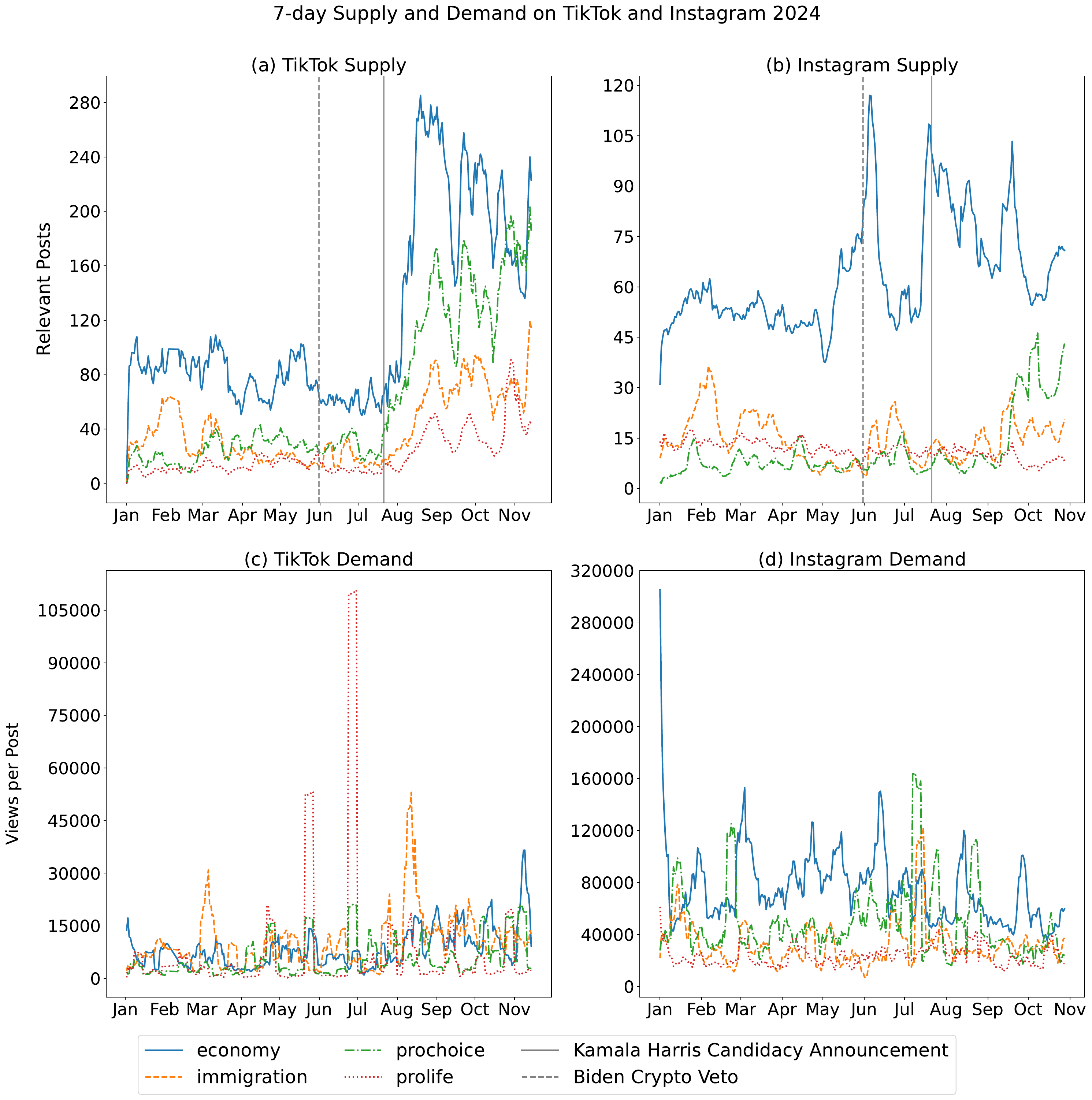}
    \caption{Supply and demand on (a, c) TikTok and (b, d) Instagram from January to November 2024 smoothed over 7-day intervals. Supply is plotted as posts per day and demand as views per post per day.}\label{fig:supply-demand}
\end{figure*}

\textbf{Supply priorities.} The supply plots in Figure~\ref{fig:supply-demand}a) and b) reflect a consistently high volume of economy-related content being created on both platforms over the 10-month period. On TikTok, 37,700 posts were classified as economy posts, compared to the second-largest category, pro-choice posts (20,371), followed by 12,518 immigration posts. Lastly, only 6,513 TikToks contained pro-life content. Instagram had a slightly different ordering of categories by post volume. Again, the economy (18,614 posts) was the largest category in terms of supply, followed by immigration (4,350), then followed by both abortion categories (3,384 pro-choice and 3,341 pro-life posts). The significant, prevailing gap between the supply of economy-related content and that of the other three categories on both platforms the weight placed on the national economy as a major issue by voters from both parties. 

Notably, Instagram displayed a far larger gap than TikTok between the supply of economy-related posts and that of posts in the other three categories. On Instagram, total economy post volume across the 10-month period was around $330\%$ higher than that of the second largest category, immigration. This percentage difference was much smaller on TikTok, with economy supply being only about $85\%$ greater than pro-choice supply. Furthermore, while abortion and immigration related content had similar levels of supply on Instagram, the supply of pro-life, pro-choice, and immigration related TikToks varied significantly. The total volume of pro-choice content was nearly double that of immigration content, and more than triple the supply of pro-life content. 

\textbf{Response to major events.} Supply trends on the platforms both saw a major shift, although across two different date intervals. On TikTok, supply across all four categories rose significantly in late July after Biden's endorsement of Harris. During a 4-week interval beginning one week before the candidacy announcement, economy-related content rose by 341\%, immigration by 256\%, pro-choice by 417\%, and lastly, pro-life oriented content by 261\%. Instagram did not see a similar response to this news, instead, a significant spike in the supply of economy related content on the platform occurred around the end of May, coinciding with developments in federal cryptocurrency regulation: the passage of FIT21 in the House of Representatives on May 22 to distinguish jurisdictional boundaries between the Securities and Exchange Commission (SEC) and Commodity Futures Trading Commission (CFTC) over the regulation of digital assets and President Biden's May 31 veto of a measure aiming to overturn SEC's crypto accounting policy, SAB 121, a controversial policy among cryptocurrency investors and financial institutions. This increase in the volume of economy-related content notably included a significant number of cryptocurrency-related posts (containing mentions of ``crypto", ``bitcoin", and ``dogecoin"). These posts constituted 0.63\% of all posts in the Instagram dataset but made up 0.81\% of posts produced in the 3-week interval between May 22 and June 12, many reflecting an anti-regulatory attitude towards the government regulation of cryptocurrency assets.

\textbf{Divergences in demand.}  While the order of content supply priorities was similar across the two platforms, major differences emerged in demand patterns: the volatility of user demand and the category of content that commands the most attention. View patterns on TikTok had a much higher degree of volatility than trends on Instagram. While the average TikTok generated less views than the average Instagram post, TikToks had more potential for virality, leading to extreme peaks in demand on the platform (Fig. 1(c)). The most viral TikToks contributing to these peaks pertained to moralized issues, namely pro-life aligned abortion and immigration. Meanwhile, Instagram lacked the same extreme outliers of high demand (Fig. 1(d)). On the latter platform, the category that garnered the most views was the economy (averaging around 10,290 views/post), a topic that is typically not placed in a moralized frame in online discourse. Demand trends on Instagram also deviate from those of TikTok in that while abortion-related content occasionally gained some amount of virality, these posts employed a pro-choice perspective. In fact, pro-life content gained the least viewers on Instagram out of all four categories, a stark contrast to the high performance of the same content on TikTok.

Table~\ref{tab:instagram} and~\ref{tab:tiktok} show the changes in volatility for Instagram and TikTok, respectively, after Biden stepped down for Harris. There are a few immediate observations. TikTok exhibits extreme demand spikes (up to 105,000 views/post) with high variance between time periods. In comparison, Instagram shows dampened, stable demand (max ~320,000 views/post but more consistent). TikTok's peaks correspond to moralized content (pro-life, immigration), while Instagram's stable high performance goes to economy content.

\begin{table}[t]
\centering
\small
\caption{Volatility in Instagram Content Before and After Harris Nomination}
\label{tab:instagram}
\begin{tabular}{lccc}
\toprule
\multicolumn{4}{c}{\textbf{Posts}} \\
\cmidrule(lr){1-4}
Topic & Before & After & $\Delta$ \\
\midrule
Economy     & 0.05 & 0.05 & -0.00 \\
Immigration & 0.12 & 0.12 & +0.00 \\
Pro-choice  & 0.13 & 0.12 & -0.01 \\
Pro-life    & 0.06 & 0.09 & +0.03 \\
\midrule
\multicolumn{4}{c}{\textbf{Views}} \\
\cmidrule(lr){1-4}
Topic & Before & After & $\Delta$ \\
\midrule
Economy     & 0.14 & 0.11 & -0.02 \\
Immigration & 0.24 & 0.13 & -0.11 \\
Pro-choice  & 0.28 & 0.25 & -0.03 \\
Pro-life    & 0.14 & 0.20 & +0.06 \\
\bottomrule
\end{tabular}
\end{table}

\begin{table}[t]
\centering
\small
\caption{Volatility in TikTok Content Before and After Harris Nomination}
\label{tab:tiktok}
\begin{tabular}{lccc}
\toprule
\multicolumn{4}{c}{\textbf{Posts}} \\
\cmidrule(lr){1-4}
Topic & Before & After & $\Delta$ \\
\midrule
Economy     & 2.86 & 0.09 & -2.77 \\
Immigration & 2.77 & 0.09 & -2.68 \\
Pro-choice  & 2.68 & 0.09 & -2.60 \\
Pro-life    & 2.73 & 0.11 & -2.62 \\
\midrule
\multicolumn{4}{c}{\textbf{Views}} \\
\cmidrule(lr){1-4}
Topic & Before & After & $\Delta$ \\
\midrule
Economy     & 0.29 & 0.27 & -0.02 \\
Immigration & 0.40 & 0.26 & -0.14 \\
Pro-choice  & 0.50 & 0.37 & -0.13 \\
Pro-life    & 0.77 & 0.57 & -0.20 \\
\bottomrule
\end{tabular}
\end{table}

Supply volatility on TikTok reduced from 2.68-2.86 to 0.09-0.11 after Harris's nomination, corresponding to a ~96\% reduction, which suggests Harris's entry transformed sporadic political posting into sustained, consistent content creation. The simultaneous volume increase indicates this wasn't a decrease in activity but a shift from irregular bursts to steady production. In comparison, Instagram volatility remained essentially unchanged (0.05-0.13 for posts, 0.11-0.28 for views). The Harris nomination had no stabilizing or destabilizing effect on Instagram content patterns. This stability reflects Instagram's network-driven model, where established accounts post consistently regardless of events. This is consistent with the literature which shows legacy and news accounts having a greater footprint on Instagram than on TikTok~\cite{changvisual}.

In sum, on TikTok, supply and demand are inversely related: economy dominates supply but moralized content (pro-life, immigration) dominates demand peaks. On Instagram, supply and demand are aligned, where economy leads both metrics. This suggests TikTok's algorithm enables "dark horse" virality, where content from smaller supply categories can still achieve massive reach. This answers \textbf{RQ1}. 

\begin{figure*}[!htb]
    \centering
    \includegraphics[width=\textwidth]{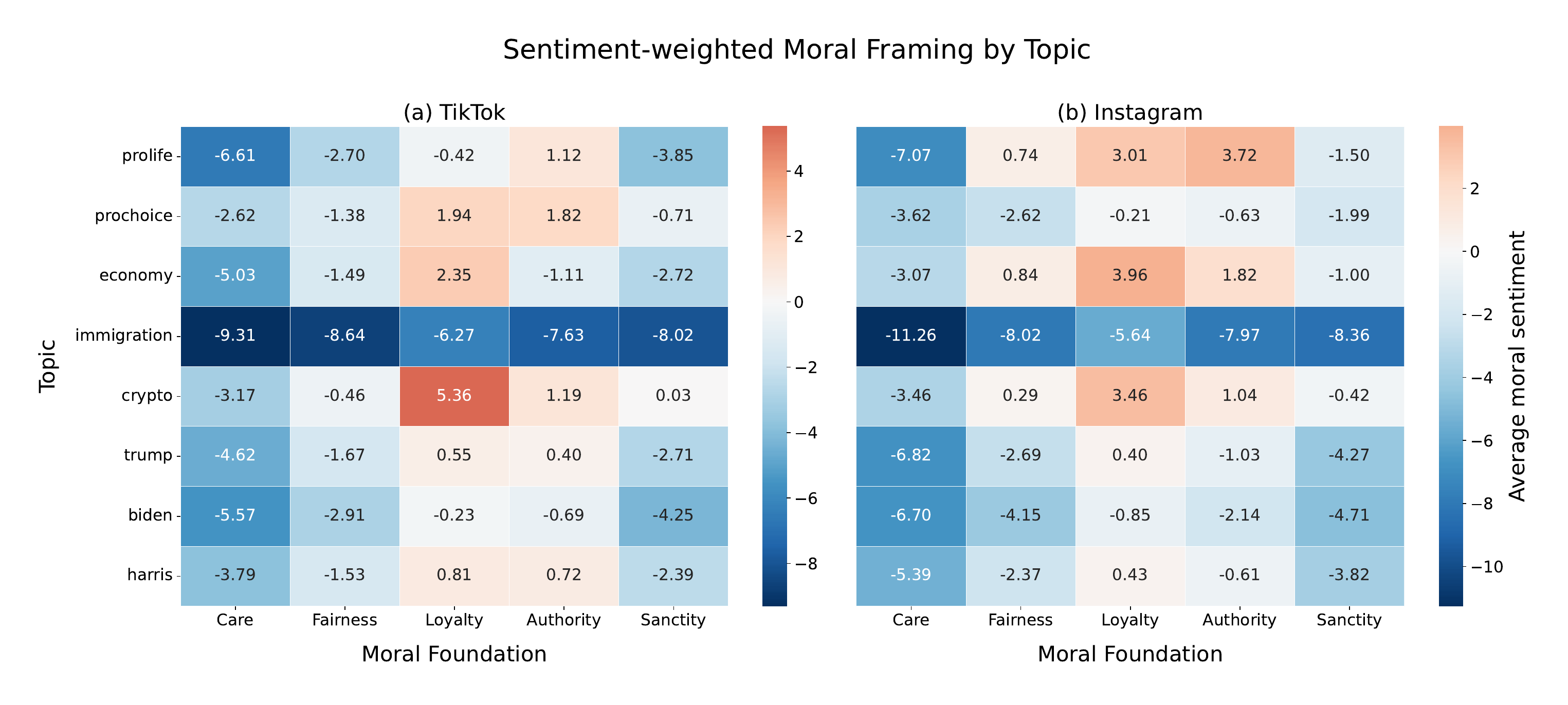}
    \caption{Platform differences in sentiment-weighted moral framing across topics and candidates. Heatmaps show average moral sentiment by foundation on TikTok (a) and Instagram (b), estimated using the extended Moral Foundations Dictionary (eMFD) with all word–foundation probabilities.}\label{fig:heatmap}
\end{figure*}

\begin{figure}[!htb]
    \centering
    \includegraphics[width=\linewidth]{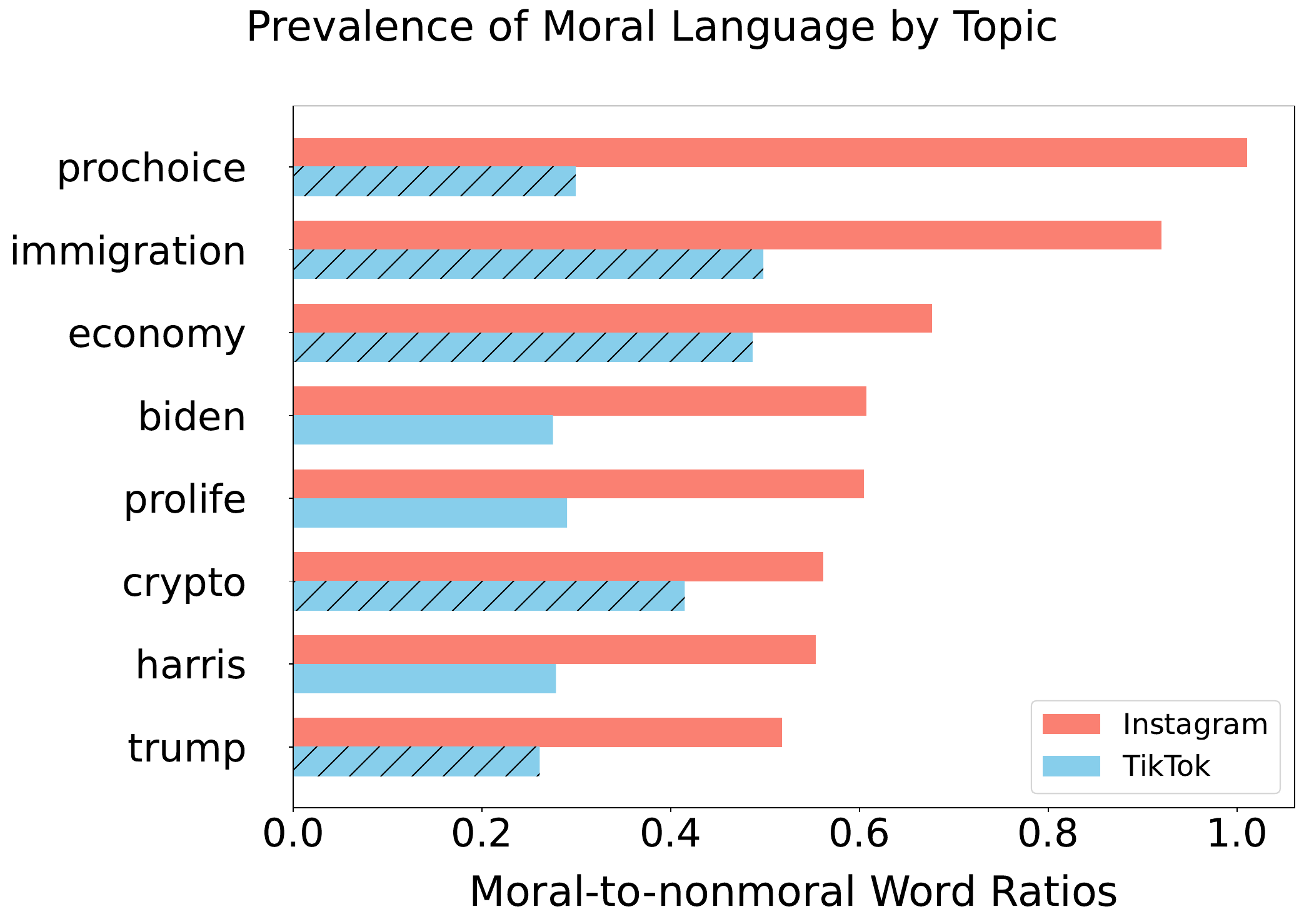}
    \caption{Average percentage of moral language across all posts, by topic.} \label{fig:total_moral}
\end{figure}

\subsection{Pragmatic Topics can be Moralized}

The temporal patterns indicate clear, issue-based divergences. What might be causing these differences? We investigate this question by considering the cross-section of key electoral issues and moral foundations. Figure~\ref{fig:heatmap} presents heatmaps showing the average moral sentiment across five moral foundations (Care, Fairness, Loyalty, Authority, Sanctity) for different topics and candidates on TikTok (2a) and Instagram (2b). The color intensity represents the strength and direction of moral framing, with darker blues indicating stronger negative sentiment (vice-related) and lighter colors indicating less negative or more positive sentiment (virtue-related). Immigration shows the most consistently negative moral framing across all foundations on both platforms (darkest blue), suggesting this topic is discussed through vice-related moral language emphasizing harm, unfairness, betrayal, subversion, and degradation. Pro-life content shows relatively less negative framing, particularly on care and sanctity dimensions, aligning with pro-life rhetoric that emphasizes protecting life (care reframed) and the sacred value of life (sanctity virtue). Pro-choice content displays moderate negative framing, stronger on care and fairness dimensions, consistent with frames about harm to women and reproductive rights violations.

Critical platform-specific divergences emerge in economy content: on TikTok, economy content scores negatively on care (-5.03) but positively on loyalty (2.35), while on Instagram it scores even more positively on loyalty (3.96) and authority (3.72). This suggests economy discourse—particularly cryptocurrency content—is moralized through binding foundations about government overreach and in-group financial freedom. Cryptocurrency specifically shows strong positive loyalty framing on both platforms (TikTok: 5.36, Instagram: 3.46), the highest loyalty score of any category, indicating crypto is discussed as an in-group identity marker and resistance to authority. Regarding candidate framing, Trump shows consistently less negative framing than Biden across most foundations on both platforms. Harris occupies a middle position, with framing similar to Trump's on loyalty and authority dimensions but more negative on care. Loyalty emerges as the key differentiator across topics: economy and crypto content invoke positive loyalty frames (in-group solidarity), while immigration invokes strong negative loyalty frames (betrayal of national community). Authority shows similar patterns, with economy/crypto content framing regulation as subversion while other topics show authority violations. Care and Fairness remain predominantly negative across topics, suggesting most political discourse emphasizes harms and injustices rather than compassion and equity.

Figure~\ref{fig:total_moral} shows the average percentage of moral language across the two platforms, by topic. Consistently, immigration, the economy, and crypto all have elevated levels of moral language. However, Instagram has significantly higher ratios (and therefore frequencies) of moral language from the supply-side.

These findings provide critical empirical support for the MAD (Motivation, Attention, Design) model of moral contagion while extending it in important ways. The divergent moralization of economic issues—particularly cryptocurrency regulation—challenges traditional assumptions about which issues naturally lend themselves to moral framing. While abortion and immigration have long been understood as inherently moralized issues in American politics, our finding that economy content invokes binding moral foundations (loyalty, authority) more strongly than individualizing foundations (care, fairness) suggests that moralization is not simply an attribute of issues themselves but emerges from issue characteristics, platform affordances, and strategic framing choices. The cryptocurrency case is particularly instructive: this issue became moralized through frames of government overreach (authority subversion) and in-group financial identity (loyalty), demonstrating how emerging policy debates can be rapidly moralized when platform dynamics favor such framing. This answers \textbf{RQ2}.

\subsection{Issue Fragmentation Reflects Platform Design}

\begin{figure}[!htb]
    \begin{subfigure}{0.5\textwidth}
        \includegraphics[width=0.85\textwidth]{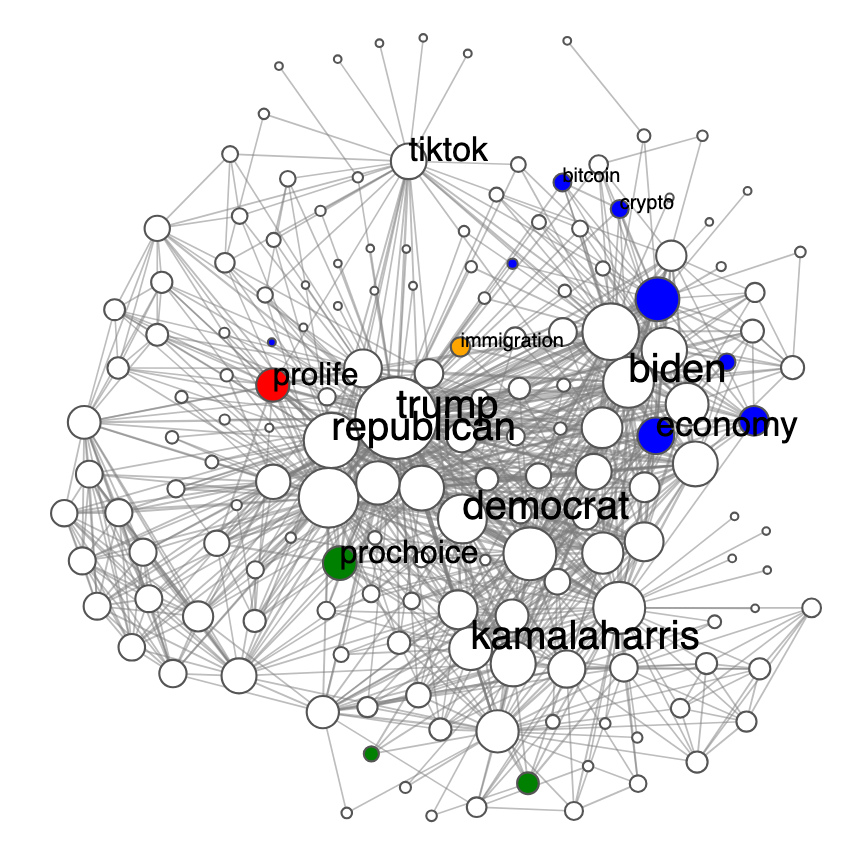}
        \caption{TikTok}
    \end{subfigure}
    \begin{subfigure}{0.5\textwidth}
        \includegraphics[width=\textwidth]{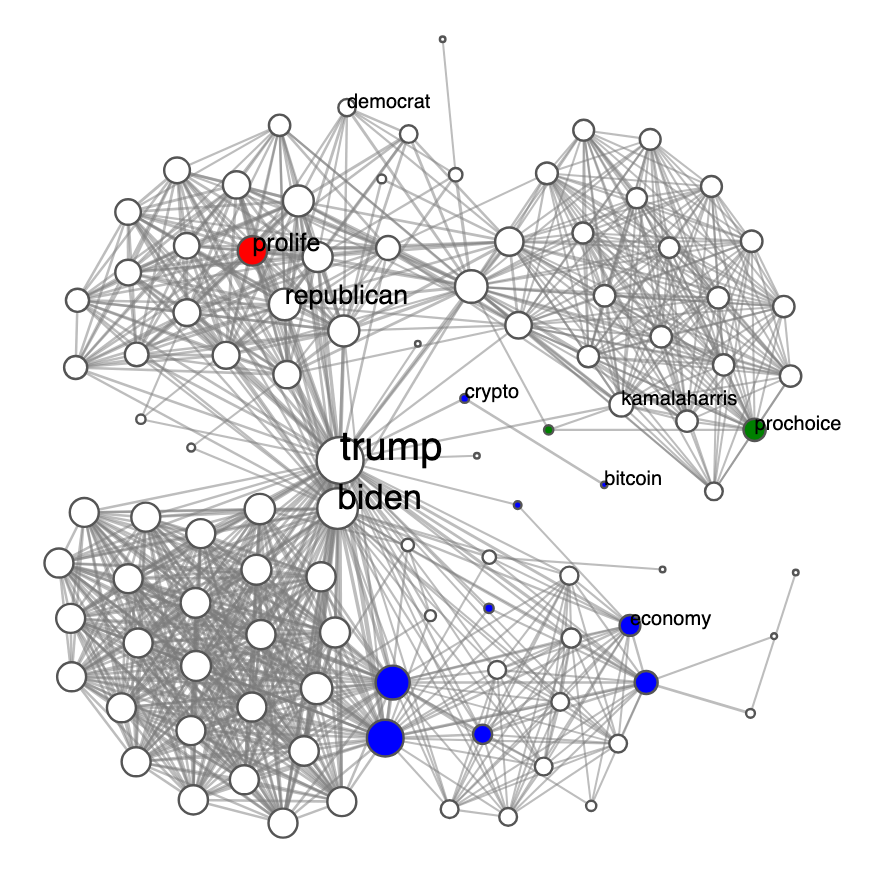}
        \caption{Instagram}
    \end{subfigure}
    \caption{Most common 1000 hashtag pairs on topically relevant (a) TikTok and (b) Instagram are visualized as networks where each node represents a hashtag and its size denotes the frequency of usage. Nodes are connected if the respective hashtags have been used together on at least one post. Hashtags categorized as economy, immigration, pro-choice, and pro-life correspond to blue, orange, green, and red nodes, respectively. Uncategorized hashtags are represented by white nodes.}
\end{figure}

The network visualizations of hashtag use on TikTok (Fig. 4(a)) and Instagram (Fig. 4(b)) provide further insight into the ideological frames present in the supply of political content on both platforms. Based on previous analysis of voter interests during the 2024 election, we expected right-leaning discourse to cluster around immigration and pro-life aligned hashtags. Similarly, we expected pro-choice hashtags to center around other left-leaning hashtags. The position of economy-related hashtags was more complicated to predict, as the issue was of importance to a majority of voters across both parties.

On both platforms, \textit{\#trump} was the most frequently used (politically relevant) hashtag in the dataset, with 437,262 TikToks and 104,501 Instagram posts featuring the hashtag. Hashtags denoting other presidential candidates were also prevalent, with 187,637 uses of \textit{\#biden} and 159,915 occurrences of \textit{\#kamalaharris} on TikTok. Comparatively on Instagram, there were 46,600 counts of \textit{\#biden} and 47,275 uses of \textit{\#kamalaharris}. Critically, Biden's node appears larger than Harris's on both networks despite similar raw frequencies on Instagram, indicating that Biden hashtags co-occurred more frequently with the four topical categories of interest. This suggests Harris remained somewhat peripheral to the core issue debates we examined—abortion, immigration, economy—which aligns with the temporal finding that Harris's campaign surge on TikTok did not translate to Instagram and with prior research showing that female candidates face exclusion from economic credibility discussions during periods of economic uncertainty~\cite{lei2018economic,chang2024will}.

These ideological clustering patterns confirm expected partisan alignments while revealing platform-specific variations in issue ownership. Pro-life hashtags cluster near Trump and Republican-aligned content on both platforms, consistent with abortion restrictions being a core Republican position. Pro-choice hashtags, conversely, cluster tightly around Harris rather than around broader Democratic or liberal identity markers. This finding is particularly striking: it suggests that abortion rights discourse in 2024 became specifically personalized around Harris's candidacy rather than articulated as a general Democratic position. This personalization likely reflects both Harris's identity as a female candidate and the strategic emphasis her campaign placed on abortion rights following the Dobbs decision. The gendered nature of this clustering has important implications—it may have strengthened Harris's appeal to voters prioritizing reproductive rights while simultaneously reinforcing gender-stereotyped issue associations that could undermine her perceived competence on other issues like the economy.


On both platforms, nodes representing President Biden were the most strongly connected to economy-related content out of all three candidates, possibly stemming from dissatisfaction with the economy during his presidential term. Nodes representing cryptocurrency related content clustered around Biden's node in the TikTok network but were more connected to Trump's node on Instagram. The TikTok pattern—crypto clustering with Biden—likely reflects content criticizing Biden's regulatory actions, particularly his veto of the SAB 121 resolution that would have overturned SEC cryptocurrency accounting requirements. This criticism frame positions Biden as the antagonist in crypto discourse, making him the focal point even for content opposing his policies. Conversely, Instagram's pattern—crypto clustering with Trump—suggests promotional or supportive framing that positions Trump as the pro-crypto candidate. This interpretation aligns with Trump's explicit appeals to cryptocurrency investors during the 2024 campaign and suggests that Instagram's user base (or its algorithmic amplification patterns) favored pro-Trump crypto advocacy over anti-Biden crypto criticism. Lastly, Harris' distance from these nodes aligns with previous findings that confidence in female candidates was undermined during periods of perceived economic uncertainty \cite{lei2018economic}. The exclusion of Harris from economy-related discourse is also a likely factor in the small size of the \textit{\#kamalaharris} node relative to the nodes of the other two candidates, given how prevalent supply of economy content across both platforms.

Finally, most critically, the two networks display notable structural differences. the most theoretically significant structural feature: TikTok's network exhibits a circular, interconnected topology while Instagram's shows a fragmented, clustered structure. TikTok's circular architecture suggests high interconnectivity across different political topics and ideological positions, where hashtags from different categories—pro-life, pro-choice, immigration, economy—maintain multiple pathways to one another through shared intermediary nodes. This structure facilitates cross-cutting exposure. In contrast, Instagram's fragmented structure, reveals distinct, semi-isolated clusters with weak bridges between them—pro-choice content clusters tightly around Harris and Trump-related content forms its own dense region. This balkanized topology reflects Instagram's hybrid algorithm where established accounts can reach followers without bridging to other communities. In sum, divergences in the semantic network reflect differences in the underlying architecture, as predicted by the MAD model.

Substantively, this fragmentation also structurally explains Harris's positioning challenge. Her close proximity to pro-choice content and distance from economy nodes physically isolated her from Instagram's largest discourse cluster, while Biden's and Trump's more central, bridge-like positions in both networks allowed their content to circulate more broadly. The cryptocurrency case exemplifies this difference. TikTok simultaneously connects multiple issues and moralization frames, while on Instagram's fragmented topology Biden drew more comparisons with Trump. Together, these answer \textbf{RQ3}.

\section{Discussion}

Our analysis of over 3 million posts across TikTok and Instagram during the 2024 presidential election reveals how platform architecture fundamentally shapes which political issues become moralized and how moral content spreads. Three key findings emerge from examining supply-demand dynamics, moral foundation invocations, and semantic network structures.

First, we demonstrate platform architectures create divergent moral ecosystems, especially by demand. The fundamental decoupling of supply and demand on TikTok versus their alignment on Instagram directly reflects algorithmic design. On TikTok, economy content dominated supply but moralized content--- abortion and immigration---dominate demand. Instagram showed the opposite: economy content led both supply and stable demand. The 96\% reduction in TikTok supply volatility after Harris's nomination demonstrates how major events transform the platform's ecosystem from sporadic bursts to sustained production, while Instagram volatility remained unchanged (0.05-0.13 for posts). This divergence strongly supports the MAD model \cite{8}: TikTok's content-driven algorithm enables moralized content to overcome supply disadvantages through attention-capturing intensity, while Instagram's network-driven model amplifies established accounts producing informational content about pragmatic issues.

Second, economic issues became moralized through candidate comparisons. This is perhaps the most theoretically interesting result derives from how cryptocurrency discourse achieved the highest loyalty scores of any category, which represents a departure from traditional economic discourse centered on efficiency and pragmatic trade-offs. The May 2024 Instagram spike following FIT21 passage and Biden's SAB 121 veto shows how cryptocurrency regulation was framed not as technical financial policy but as government overreach threatening financial freedom. This challenges assumptions about which issues are ``inherently" moralized, demonstrating instead that moralization emerges from interactions between issue characteristics, platform affordances, and strategic choices. While Instagram showed significantly higher moral language ratios across all topics despite TikTok's reputation as more ``affective" suggests platform demographics and content norms matter as much as algorithmic design. Further research should address the more auditory nature of TikTok may contribute to this discrepancy, such as transcript analysis. 

Lastly, network topology reflects the level of discourse integration and fragmentation. TikTok's circular semantic network reflects the multiple pathways for cross-cutting issue exposure, while Instagram's fragmented structure shows semi-isolated clusters with weak bridges. This structural difference has concrete political consequences: Harris's tight clustering with pro-choice content and distance from economy nodes physically isolated her from Instagram's largest discourse cluster, reinforcing research on gender and electability, in prior research \cite{lei2018economic, chang2024will}. In other words, platforms amplify the "locker-room" topics mainly owned by men.

\subsection{Limitations and Future Directions}

Our study contains a few limitations. First, our hashtag-based categorization cannot fully capture internal ideological diversity within categories, the demand, or the moral framing embedded in visual and audio content. Since eMFD analyzes only linguistic components, future work should assess how images and videos invoke moral foundations. Future work should incorporate computer vision and audio analysis to assess multimodal moral framing, use supervised learning on manually coded samples to reveal within-category variation, and analyze engagement type breakdowns (likes, shares, comments) to determine whether different moral foundations trigger different sharing behaviors. 

As an observational analysis, we cannot establish causal relationships from observational data, despite  divergent responses to identical political events strongly suggest platform design shapes content dynamics, confounding factors remain in user demographics and content creator norms. Experimental manipulations varying recommendation algorithms could more definitively establish causality. Despite these limitations, our work demonstrates that as political communication shifts from text-based to visual-first platforms, understanding how algorithmic design, network topology, and moral foundation deployment interact to determine which issues get moralized becomes essential for both political practice and democratic theory.

\bibliography{bibfile}

\end{document}